\documentclass[12pt]{article}
\usepackage{graphicx}

\newcommand\pubnumber{}
\newcommand\pubdate{\today}

\def\institute{Centre for Cosmology, Particle Physics and Phenomenology\\
Universit\'e catholique de Louvain, B-1348 Louvain-la-Neuve, BELGIUM}
\def\support{\footnote{Work supported by Fonds de la Recherche Scientifique (FNRS), Belgium.}}

\def\Title#1{\begin{center} {\Large #1 } \end{center}}
\def\Author#1{\begin{center}{ \sc #1} \end{center}}
\def\Address#1{\begin{center}{ \it #1} \end{center}}

\newcommand\pubblock{\rightline{\begin{tabular}{l} \pubnumber\\
         \pubdate  \end{tabular}}}
\newenvironment{Abstract}{\begin{quotation}  }{\end{quotation}}
\newenvironment{Presented}{\begin{quotation} \begin{center} 
             PRESENTED AT\end{center}\bigskip 
      \begin{center}\begin{large}}{\end{large}\end{center} \end{quotation}}

\def\beq{\begin{equation}}
\def\eeq#1{\label{#1}\end{equation}}
\def\eeqn{\end{equation}}

\def\beqa{\begin{eqnarray}}
\def\eeqa#1{\label{#1}\end{eqnarray}}
\def\eeqan{\end{eqnarray}}

\let\bar=\overbar

\def\Dslash{\not{\hbox{\kern-4pt $D$}}}
\def\dslash{\not{\hbox{\kern-2pt $\del$}}}

\def\msb{{\bar{\ssstyle M \kern -1pt S}}}

\usepackage{xspace}

\newcommand{\wjets}{\ensuremath{\textrm{W+jets}}\xspace}

\newcommand{\ttbar}{\ensuremath{\mathrm{t}\bar{\mathrm{t}}}\xspace}

\newcommand{\pt}{\ensuremath{p_\mathrm{T}}\xspace}

\newcommand{\mtop}{\ensuremath{m_{\mu\nu \mathrm{b}}}\xspace}

\newcommand{\mtw}{\ensuremath{m_{\mathrm{T}}(\mathrm{W})}\xspace}
\newcommand{\pvmiss}{\ensuremath{\vec{p}\hspace{-0.5em}/\kern 0.5em}\xspace}

\def\colvecnext#1{
    #1
    \global\advance\colveccount-1
    \ifnum\colveccount>0
        \\
        \expandafter\colvecnext
    \else
        \end{pmatrix}
    \fi
}

\begin{document}
\begin{titlepage}
\pubblock

\vfill
\Title{Measurement of differential cross sections for t-channel single-top-quark production at 13 TeV}
\vfill
\Author{ Matthias Komm\support, on behalf of the CMS collaboration}
\Address{\institute}
\vfill
\begin{Abstract}
The production of single top quarks is a cornerstone in understanding the nature of the heaviest known elementary particle and its involvement in electroweak interactions. An early differential cross section measurement of t-channel single-top-quark production is presented. Proton-Proton collision data at a center-of-mass energy of $13~\mathrm{TeV}$ collected in 2015 were analyzed, corresponding to $2.3~\mathrm{fb}^{-1}$. The amount of signal events as a function of the top quark transverse momentum and rapidity is estimated using a multivariate discriminant. The results are unfolded to parton level and compared to predictions by various Monte-Carlo generators.
\end{Abstract}
\vfill
\begin{Presented}
$9^{th}$ International Workshop on Top Quark Physics\\
Olomouc, Czech Republic,  September 19--23, 2016
\end{Presented}
\vfill
\end{titlepage}
\def\thefootnote{\fnsymbol{footnote}}
\setcounter{footnote}{0}

\section{Introduction}

Cross section measurements of single top quark production allow to study electroweak interactions involving heavy quarks. For example, precise inclusive cross section measurements can yield model-independent limits on the CKM matrix element, $|\mathrm{V}_\mathrm{tb}|$, whereas differential measurements can test theoretical calculations and predictions by generators in great detail. Inclusive and differential t-channel single-top-quark measurements at 8~TeV can be found in Refs.~\cite{Vtb,TOP-14-004}. A first inclusive cross section measurement at 13~TeV is detailed in Ref.~\cite{TOP-16-003}.

In this note, a first differential cross section of t-channel single-top-quark production as a function of the top quark \pt and rapidity, $y=\frac{1}{2}\ln[(E+p_{z})/(E-p_{z})]$, is summarized, using the first pp collision data at $\sqrt{s}=13~\mathrm{TeV}$ recorded by the CMS experiment in 2015 corresponding to $2.3~\mathrm{fb}^{-1}$. More details can be found in Ref.~\cite{TOP-16-004}.

The measurement focuses on single top quarks decaying into a b jet, a muon, and a neutrino. The t-channel production signature consists furthermore of a light~(u/d/s), spectator quark which is characteristically scattered into the forward detector region. Muon candidates are required to be well-isolated with a transverse momentum of $\pt>22~\mathrm{GeV}$ within a pseudorapidity range of $|\eta|<2.4$. Jets with $\pt>40~\mathrm{GeV}$ are selected within $|\eta|<4.7$. A b-tagging algorithm is employed for jets within $|\eta|<2.4$ based on a multivariate discriminant to identify jets originating from hadronizing b quarks. A selection on the transverse W boson mass of $\mtw=|\vec{p}_{\mu}+\pvmiss|_\mathrm{T}>50~\mathrm{GeV}$ is applied to reduce the contamination of multijet events. In addition to the signal region~(2~jets, 1~b-tag), events from two control regions containing 3~jets with 1~or 2~b-tags are considered. These are enriched in \ttbar events which is the main background of this measurement.

\section{Signal extraction}

A neutrino candidate is found by solving for the unknown $p_{z}^{\nu}$ momentum using a W boson mass constraint, $m_{W}^{2}=(p_{\mu}+p_{\nu})^{2}$. The 4-momentum of a top quark candidate can then be reconstructed from the reconstructed decay products.

\begin{figure}[!htbp]
\begin{center}
\includegraphics[width=0.45\textwidth]{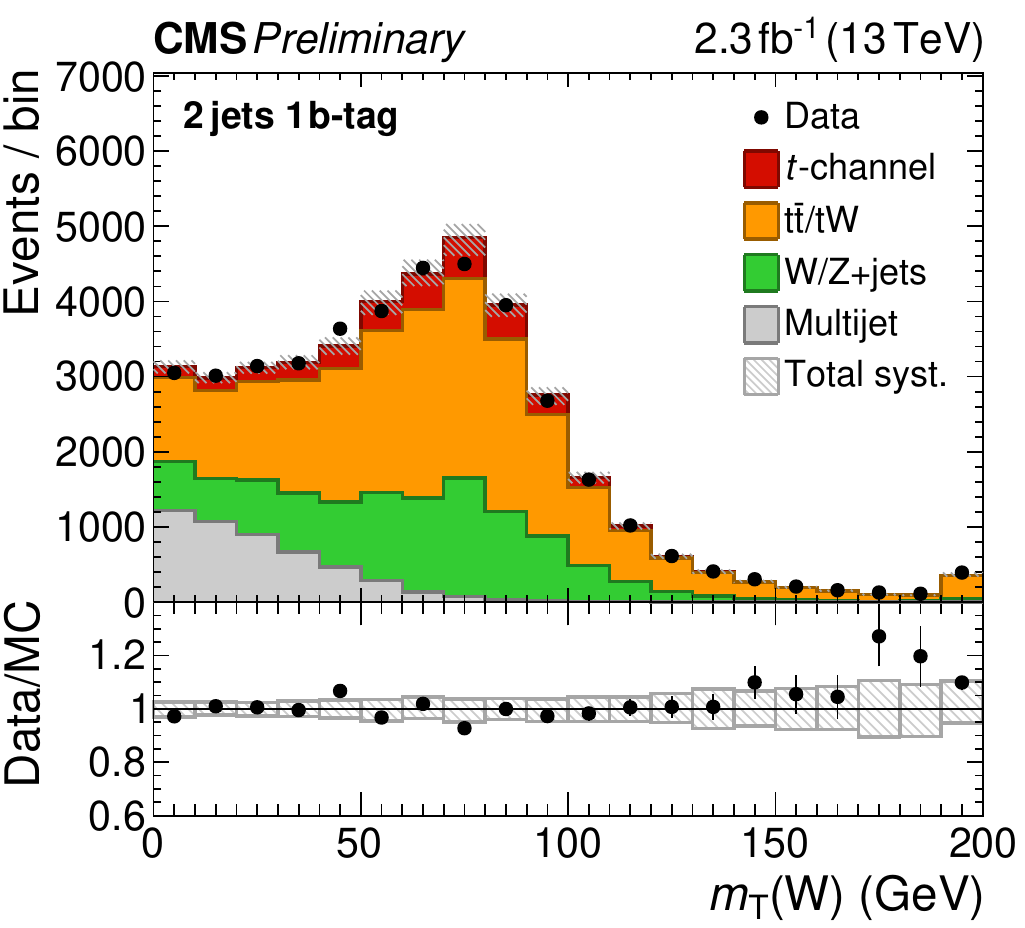}\hspace{0.05\textwidth}
\includegraphics[width=0.45\textwidth]{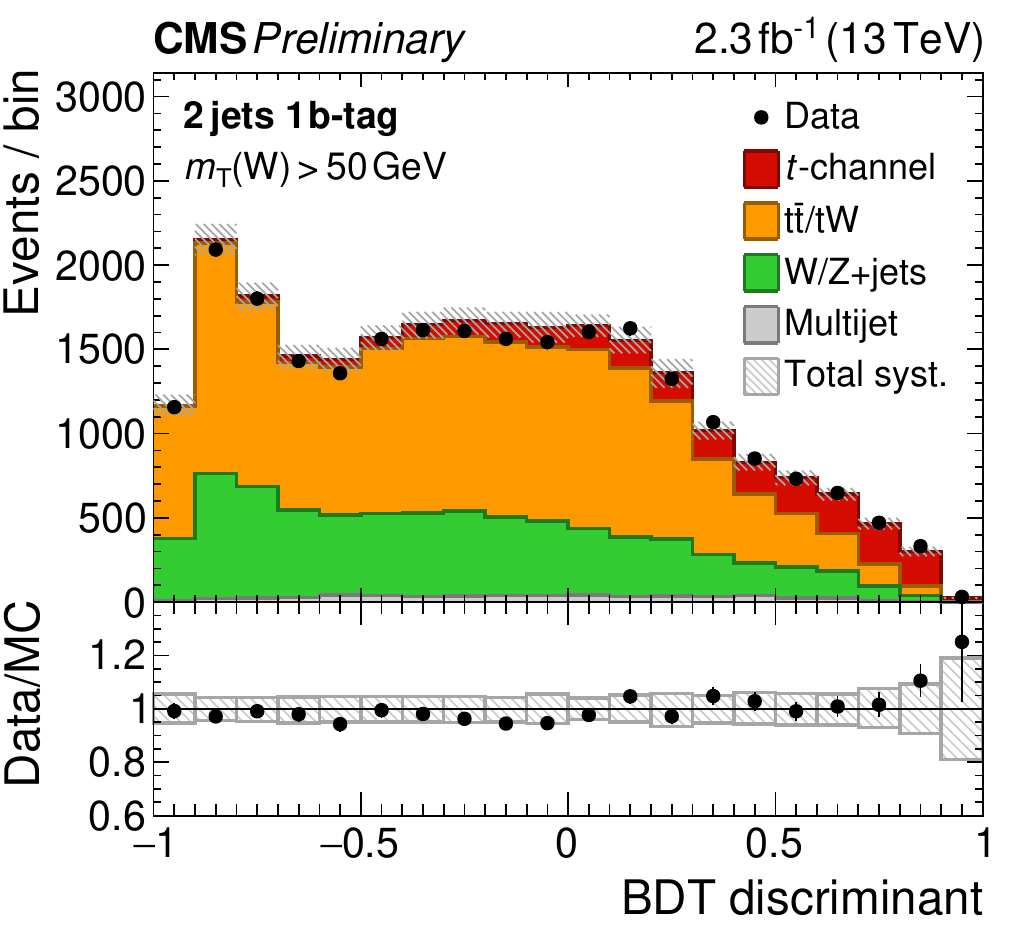}
\end{center}

\caption{\label{fig:mtwbdt}Distributions of (left)~the transverse W boson mass and (right)~the BDT discriminant while requiring $\mtw>50~\mathrm{GeV}$.}
\end{figure}

To extract the signal yield as a function of the reconstructed top quark \pt and $|y|$, multiple, exclusive maximum likelihood fits are performed in bins of the measurement. The fits utilize an extended likelihood using the $\mtw$ shape for events with $\mtw<50~\mathrm{GeV}$ and the shape of a Boosted Decision Tree~(BDT) discriminant otherwise. The BDT is trained to discriminate signal against \ttbar and \wjets events by exploiting the following event observables: $\mtop$, $|\eta_\mathrm{light~jet}|$, $\Delta R(\mathrm{b\mbox{-}jet,~light~jet})$, $|\Delta\eta(\mathrm{b\mbox{-}jet,\mu})|$, and \mtw. The low $\mtw$ region is particularly sensitive to the multijet contribution whose shape is estimated from data using a sideband with inverted muon isolation. Figure~\ref{fig:mtwbdt} shows the $\mtw$ and BDT distributions after scaling the predictions to the inclusive fit result. A good agreement between data and predictions is observed.

\begin{figure}[!htbp]
\begin{center}
\includegraphics[width=0.43\textwidth]{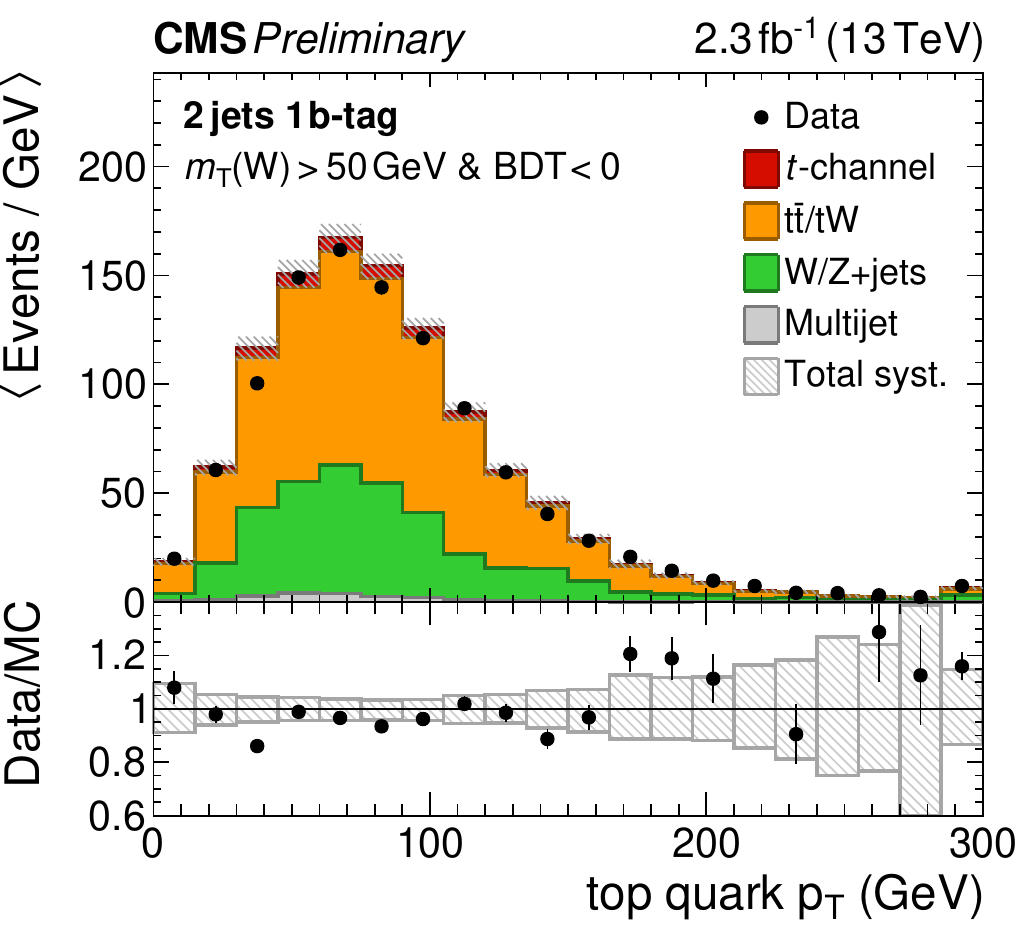}\hspace{0.05\textwidth}
\includegraphics[width=0.43\textwidth]{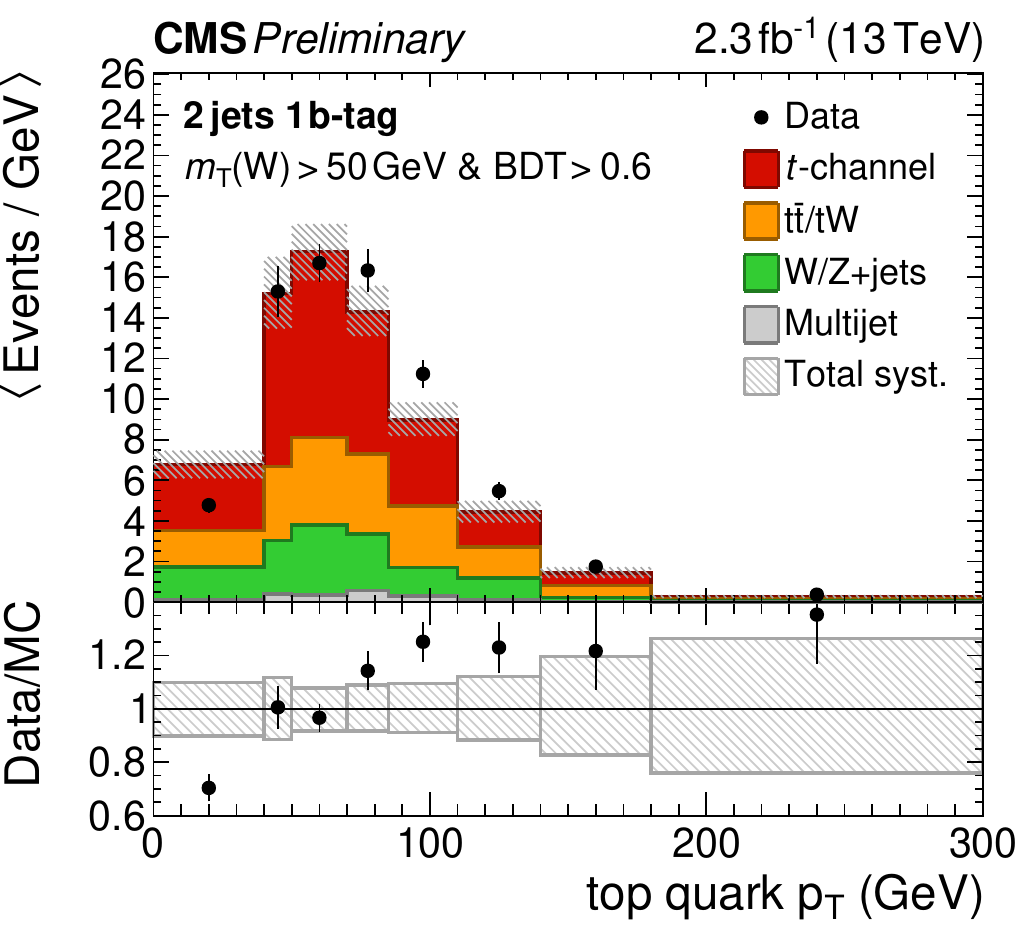}
\end{center}

\caption{\label{fig:recotop-pt}Distributions of the top quark transverse momentum for events passing $\mtw>50~\mathrm{GeV}$: (left)~signal-depleted phase space, $\mathrm{BDT}<0$, and (right)~signal-enhanced phase space, $\mathrm{BDT}>0.6$. Predictions are normalized to the inclusive fit result.}
\end{figure}

\begin{figure}[!htbp]
\begin{center}
\includegraphics[width=0.43\textwidth]{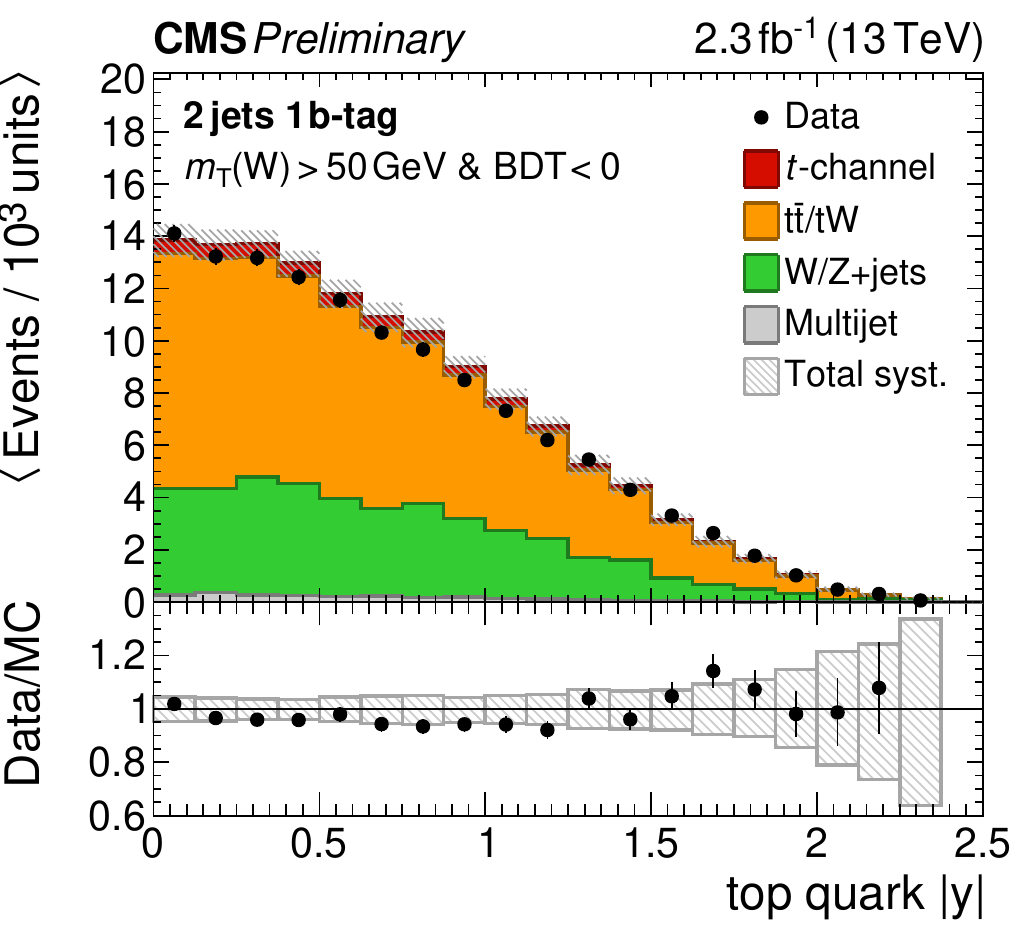}\hspace{0.05\textwidth}
\includegraphics[width=0.43\textwidth]{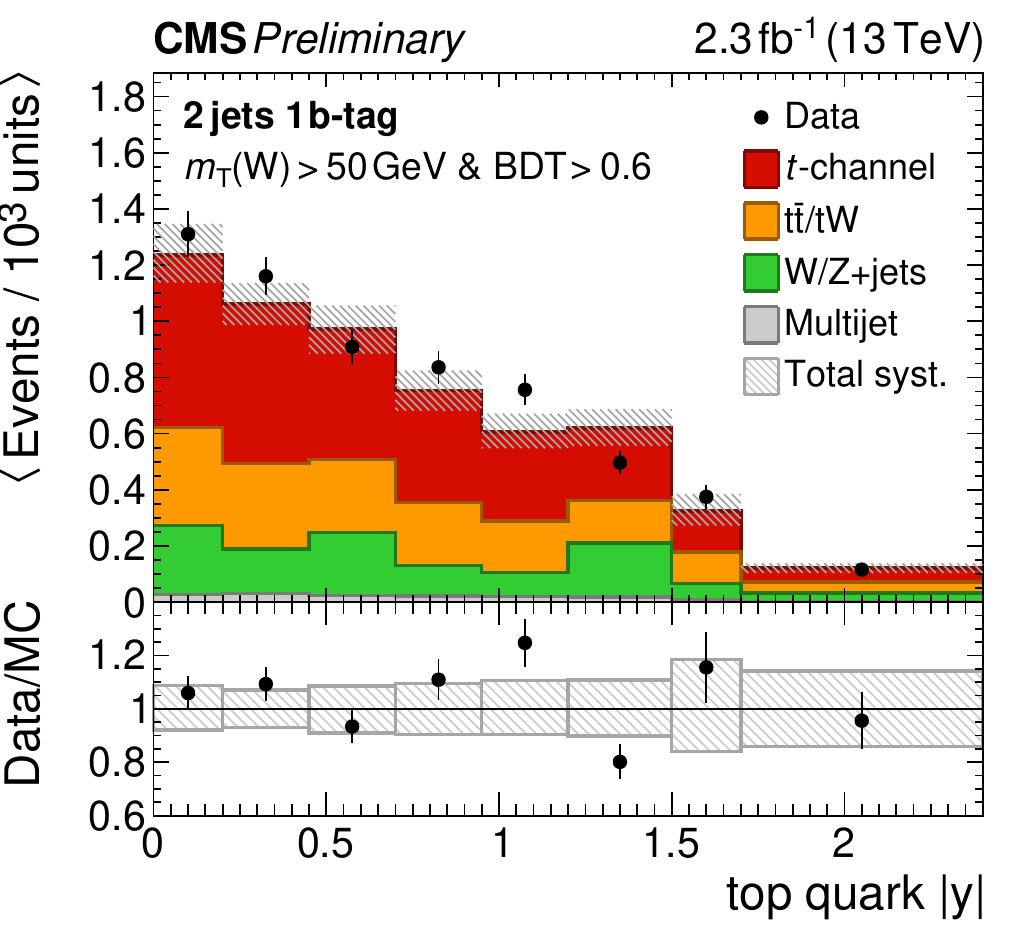}
\end{center}

\caption{\label{fig:recotop-y}Distributions of the top quark rapidity for events passing $\mtw>50~\mathrm{GeV}$: (left)~signal-depleted phase space, $\mathrm{BDT}<0$, and (right)~signal-enhanced phase space, $\mathrm{BDT}>0.6$. Predictions are normalized to the inclusive fit result.}
\end{figure}

The distributions of the top quark \pt and $|y|$ at reconstruction level are shown in Figs.~\ref{fig:recotop-pt} and~\ref{fig:recotop-y}. A signal-depleted (signal-enhanced) phase space defined by $\mtw>50~\mathrm{GeV}$ and $\mathrm{BDT}<0$ ($\mathrm{BDT}>0.6$) is shown on the left (right), respectively. A good agreement is found for the background predictions in the signal-depleted phase space for both observables. In the signal-enhanced phase space, the distribution of the top quark rapidity agrees well with the prediction. However, data is displaying a slightly harder top quark \pt spectrum compared to the prediction.

To infer the corresponding parton-level differential cross sections, the exclusive fit results are passed to an unfolding procedure. Regularized unfolding based on the curvature of the unfolded spectrum is employed. 

The measurement is affected by various sources of systematic uncertainties. New templates for signal and background processes are derived reflecting the impact of each uncertainty source on the predictions. For each systematic variation, the fit and unfolding procedures are repeated using the new templates. The cross sections and their uncertainties are taken from the combination of all variations per bin. The individual uncertainty sources with the largest impact on the measurement are found to be the data statistics~(10\%-25\%), the choice of renormalization and factorization scale of the signal process~(10\%-15\%), the top-quark mass~(10\%-20\%), and the jet energy corrections~(10\%-15\%).

\section{Results}

\begin{figure}[!htbp]
\begin{center}
\includegraphics[width=0.43\textwidth]{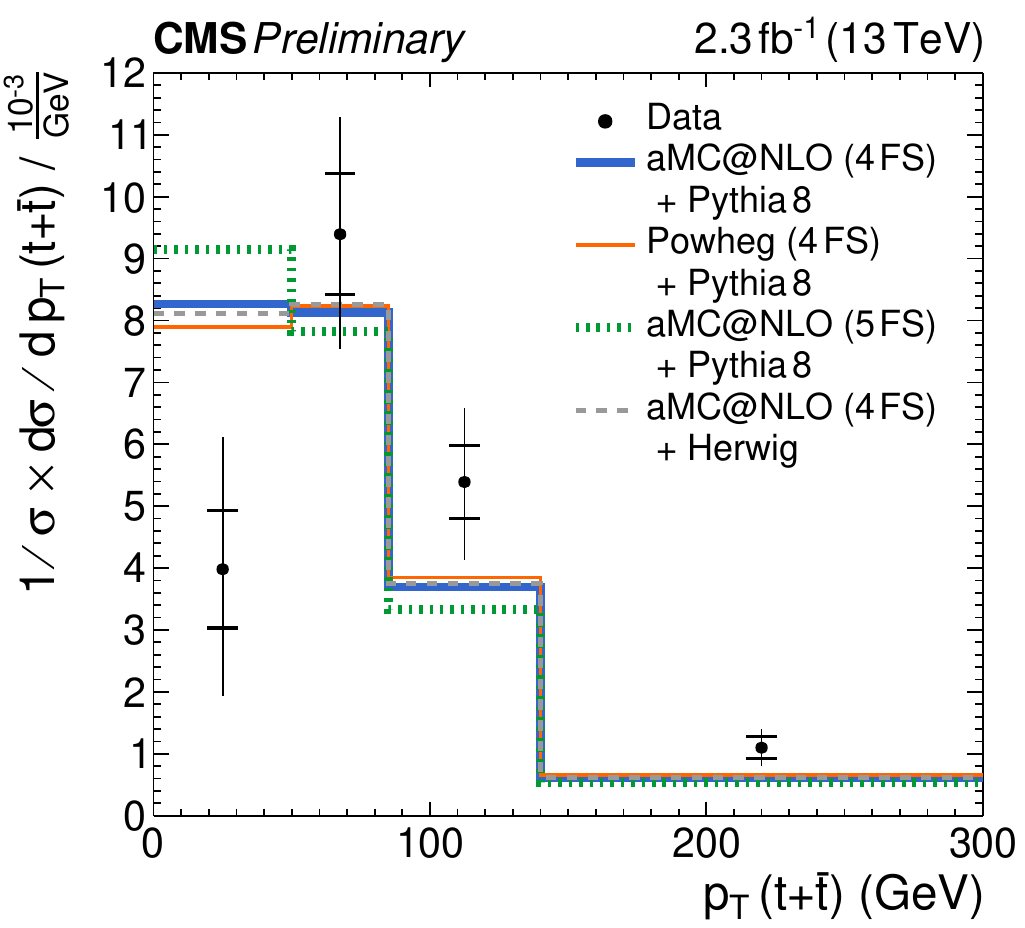}\hspace{0.05\textwidth}
\includegraphics[width=0.43\textwidth]{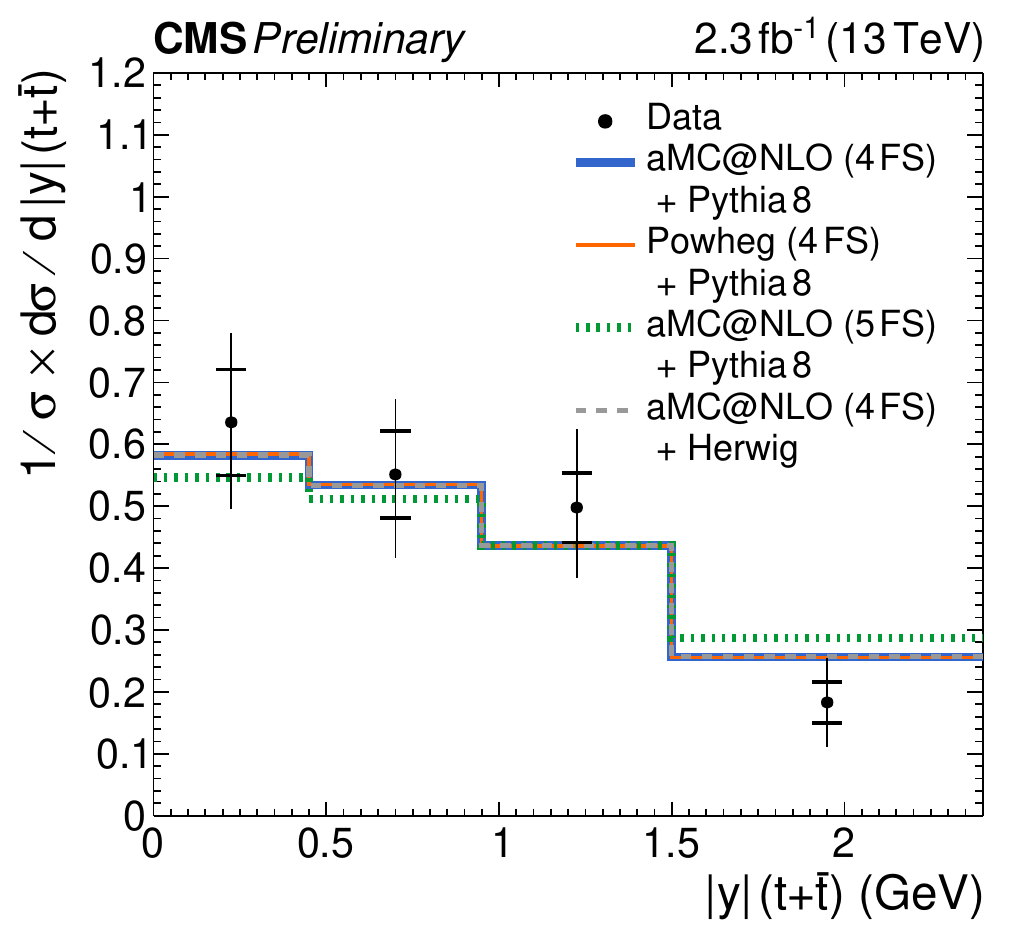}
\end{center}

\caption{\label{fig:result}Normalized differential t-channel single-top-quark cross section as a function of the parton-level top quark (left)~transverse momentum and (right)~rapidity.}
\end{figure}

The measured differential cross section of t-channel single-top-quark production are shown in Fig.~\ref{fig:result} normalized to the inclusive cross section. The results are compared to the predictions by various generators~(a\textsc{mc@nlo}, \textsc{powheg}), parton showers~(\textsc{pythia}8, \textsc{herwig}), and flavour-scheme combinations. The unfolded data is found in agreement with the predictions within the uncertainties.

\section{Conclusion}

A first measurement of t-channel single-top-quark differential cross section as a function of the top quark transverse momentum and rapidity has been performed using pp collision data at $\sqrt{s}=13~\mathrm{TeV}$ corresponding to $2.3~\mathrm{fb}^{-1}$. The results are compared to the predictions by various generators. No significant deviation is observed.

\end{document}